# The Energy Deposition on the ILC Realistic Undulator Wall

Alharbi.K[1,2,3], Alrashdi. A[3], Riemann. S[2]

[1] Hamburg University, Germany

[2] DESY-Zeuthen, Germany

[3] King Abdul-Aziz City for Science and Technology, Riyadh, Kingdom of Saudi Arabia

## Abstract

Since the undulator wall is being bombarded by photon produced in the ILC helical undulator, masks were installed inside the undulator to protect the superconducting undulator as well as the vacuum. The photon energy spectrum was used to calculate the incident power. HUSR software was used to simulate the photon energy spectrum per meter inside the undulator. The influence of adding masks inside the undulator on the photon polarisation and energy spectrum was also studied.

## Introduction

According to (Flottmann, 1993), circularly polarized photons, with an energy of several MeV, are produced when an electron beam, containing hundreds of GeV energy, passes through a helical undulator as predicted by the baseline positron source for the International Linear Collider (ILC). These photons then strike on a spinning Ti6Al4V target with 0.4 radiation lengths producing longitudinally polarized positrons through the pair production mechanism.

The parameters of the ILC positron source are adjusted accordingly to start the ILC operation at a center-of-mass energy of 250 Giga Electron Volt (GEV). Notably, 128 GeV electron beam will be sent through the helical undulator measuring 264 m in length and must be targeted at the 20 cm thickness. So the undulator system consists of 132 undulator modules, with each being 1.75 m long. Their total length of 264m produces photons with an average energy of 7.5 MeV. However, the power incident on the walls of the superconducting undulator yield heating problems due to the long undulator and the larger opening angle of the photon beam.

Thus, one possible solution for protecting the wall is to install masks between the undulator modules [2]. As a result, the energy deposition in the undulator wall could be decreased and absorbed by masks. There are other benefits for adding masks between the undulator modules. One of the most important advantages is positron polarisation. Since the photon polarisation created by a helical undulator differs with emission angle, the photon polarisation could be increased with the masks [2].

The simulation used in this study is called Helical Undulator Synchrotron Radiation (HUSR). It was developed by David Newton at the Cockcroft Institute in the UK. HUSR can simulate the photon spectrum produced by an electron beam that passes through a helical field map. The calculations in HUSR are based on equation 25 from the Kincaid paper on helical undulator radiation [3].

In the paper here, the energy deposition on the realistic undulator wall was primarily considered. Also, the effect of adding masks inside the undulator on energy deposition and photon polarisation was taken into consideration.

**Parameter Used**

The parameters used in these simulations, which are shown in table 1, are the ILC TDR positron source parameters, except the electron drive beam energy, which is 128 GeV. Additionally, the Undulator Effective Length and Undulator Deflection Parameter (K) are 264m and 0.85 respectively.

*Table 1: Undulator parameters used to calculate the photon energy spectrum produced from helical undulator.*

| Parameters | |
|---|---|
| Electron Energy (GeV) | 128 |
| Bunch population | $2 \times 10^{10}$ |
| Number of bunches | 1312 |
| Repetition rate (Hz) | 5 |
| Undulator module Length (m) | 1.75 |
| Number of undulator module | 132 |
| Net length of undulator (m) | 231 |
| Active Undulator length (m) | 264 |
| Period (mm) | 11.5 |
| K value | 0.85 |
| Magnetic field (T) | 0.79 |

## Realistic (Non-Ideal) Magnetic Field Map

Because it is impossible to build an ideal helical undulator, we need to calculate the realistic photon energy spectra produced from realistic (non-ideal) undulator. HUSR can do that when errors are added to the helical undulator magnetic field. First, a *Hall* probe was used to measure the magnetic field maps from the two helical undulator prototypes built at Rutherford Appleton Laboratory in the UK. This measurement gave an error margin of ±10% of the measured magnetic field from these two prototypes. These errors were randomly introduced into the period and field when realistic (non-Ideal) magnetic field maps were built [4].

## Photon Flux:

It is important to note that photons produced from the helical undulator give a cone of radiation and the maximum photon opening angle inside the undulator will be at the end of the undulator [2]. The photon flux produced from the 1st undulator module, measuring precisely 1.75m long, when single electron with 128 GeV energy passes through was simulated. Figure 1 shows the sketch of the undulator with the target.

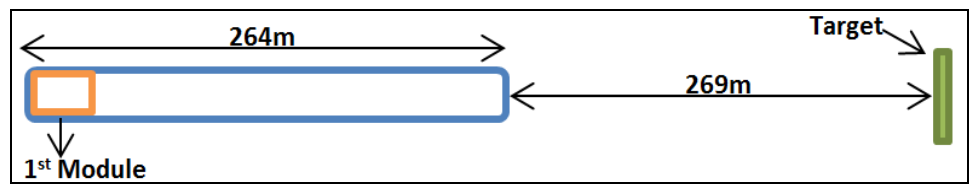

*Figure 1: the sketch of the undulator with the target.*

The flux distribution produced by the 1st module was studied at the different distance between the 1st undulator module and target. Flux was normalized to 1.

Figure 2 and 3 shows the ideal and non-ideal photon flux distributions produced by the 1st module at the middle and end of the undulator as well as at the target layer respectively, primarily taken from the left to the right.

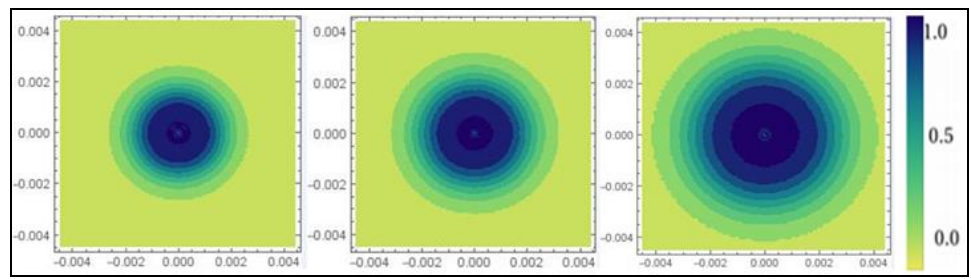

*Figure 2: Photon Flux produced by the 1st ideal undulator module at the middle and end of the undulator, and at the target layer from left to right respectively.*

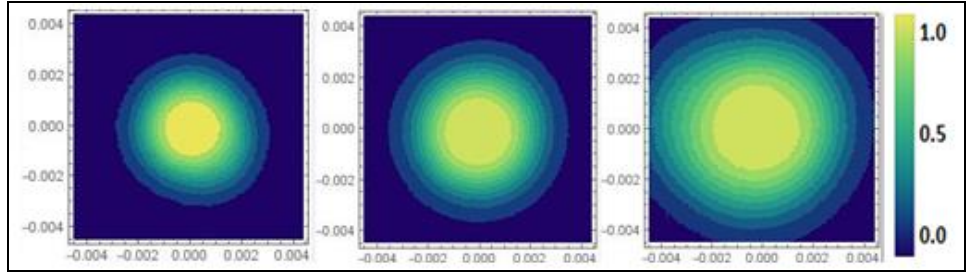

*Figure 3: Photon Flux produced by the 1ˢᵗ realistic undulator module at the middle and end of the undulator, and at the target layer from left to right respectively.*

It is clear that the radiation created by the ideal helical undulator is circularly symmetric as in figure 2. In contrast, figure 3 shows that the radiation is no longer circularly symmetric when using the non-ideal magnetic field. Additionally, since the radiation produced is as cone-shaped, there is no doubt that the radiation would hit the undulator wall when its radius becomes more significant than the undulator wall. It means that the maximum photon opening angle inside the undulator would be with photons produced by the 1ˢᵗ module. Therefore, it is expected that the 1ˢᵗ undulator module would cause the highest amount of energy deposition in the undulator wall, while the lowest amount would be generated by the last (132ⁿᵈ) undulator module.

**Energy Deposition inside the Undulator:**

Since Synchrotron radiation created by the helical undulator has an angular divergence and this radiation travels through a very long undulator, the radius of the Synchrotron radiation cross-section will become more substantial than the radius of the undulator. As a result, this radiation will hit the undulator wall and causes some quenching danger.

HUSR was used to calculate the photon spectrum produced per meter along the undulator and then the spectrum was used to calculate the incident power per meter. It was found that the accurate calculation of spectrum from HUSR was very weak when the distance between the module and observer was below 86m. Therefore, the incident powers caused by the 1ˢᵗ, 16ᵗʰ and 33ʳᵈ modules were calculated from the distance 88m, 120m, and 152m respectively, to the end of the undulator. Figure 4 shows the sketch of the undulator modules chosen to calculate the incident power per meter on the undulator wall.

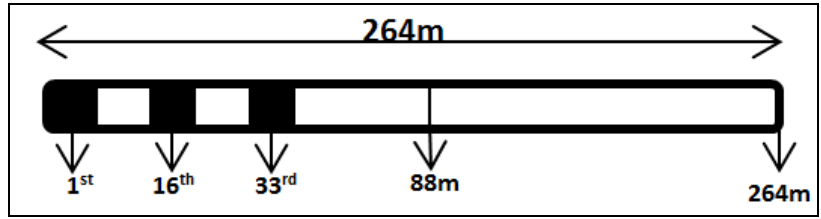

*Figure 4: The sketch of the undulator modules chosen (the 1ˢᵗ, 16ᵗʰ, and the 33ʳᵈ modules) to calculate the incident per meter along the undulator*

Since HUSR did not calculate the incident power directly, photon spectrums with different observer apertures were simulated to investigate the incident power per meter. Therefore, series of simulations with different non-ideal magnetic field maps were carried out to investigate the incident power per meter from these three modules. It was found that besides the errors on the magnetic field maps; HUSR could provide more accurate calculation when the distance between the module and observer was increased.

Since the undulator is very long, we need too much time to investigate these errors per meter. Therefore, some points were chosen to study the errors. These points were chosen by steps with 16 m width. Numbers of simulations were carried out at these locations from 88m to the end of the undulator for the 1ˢᵗ module, 120m to the end of the undulator for the 16ᵗʰ module, as well as from 152m to the end of the undulator for the 33ʳᵈ module. Figure 5 shows, PRELIMINARY, the incident power per meter caused by the 1ˢᵗ, 16ᵗʰand 33ʳᵈ undulator modules.

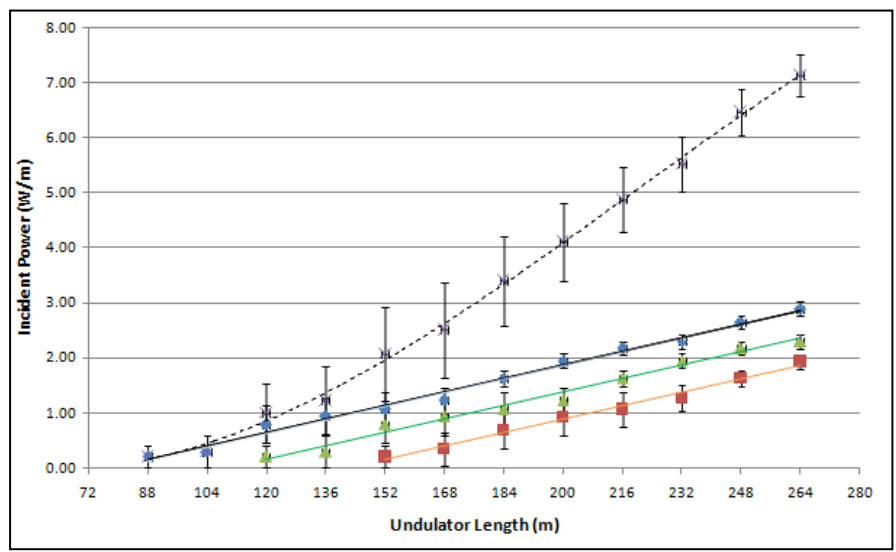

*Figure 5: Shows the incident power per meter in the undulator wall (PRELIMINARY).*

The black line shows the incident power caused by the 1st module and the green line presents the incident power from the 16th undulator module. The orange line presents the incident power from the 33rd undulator module. The dashed line gives the total energy deposition caused by all the three undulator modules.

It was found that the errors at the end of the undulator are much lower than that at 88m. The errors at a small distance between the module and the observer could reach to 90%, such as the errors at 88m. In contrast, the calculation error at the end of the undulator was about 6%.

Because 1 W/m is the maximum allowable heat load in the Superconducting undulator (2), it is clear that this level could be reachable by the 1st module which deposited about 2.9 ±0.3 W/m at the end of the undulator. Also, the total incident power caused by these modules at the end of the undulator could reach 7.14 ±0.15 W/m. Therefore, to absorb some of this energy a collimation system is needed.

**Masks**

It was necessary to have the undulator fitted with masks containing apertures smaller than the opening of the helical undulator itself to protect its wall. The mask aperture and distance between masks were estimated using a 150 GeV electron beam in the previous study [5]. In this particular study, a 128 GeV electron beam was used, with 4.4 mm aperture masks placed inside the undulator. The distance between masks was 16m. Figure 6 shows the sketch of the modules between two masks.

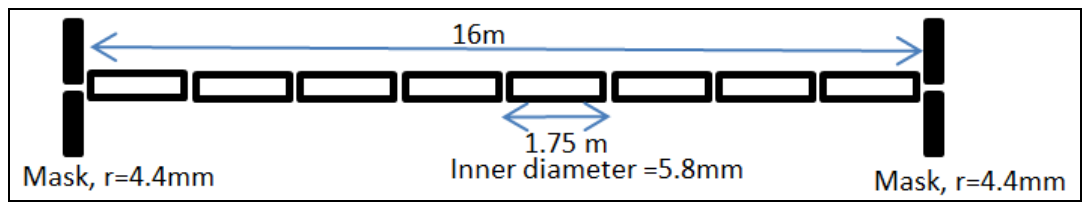

*Figure 6: the sketch of the modules between two masks*

HUSR was used to investigate the incident power per meter, primarily utilising the photon spectrum. The calculation errors were high. Therefore, several simulations with different nun-ideal magnetic field maps were carried out to investigate the incident power between two masks. Figure 7 below shows the preliminarily incident power per meter between two masks. It is clear that the incident power on the wall would be absorbed by the masks. However, there were high errors.

The finding that adding a mask inside the undulator can affect the photon spectrum and polarisation was significant in studying the influence of photon energy spectrums as well as polarisation.

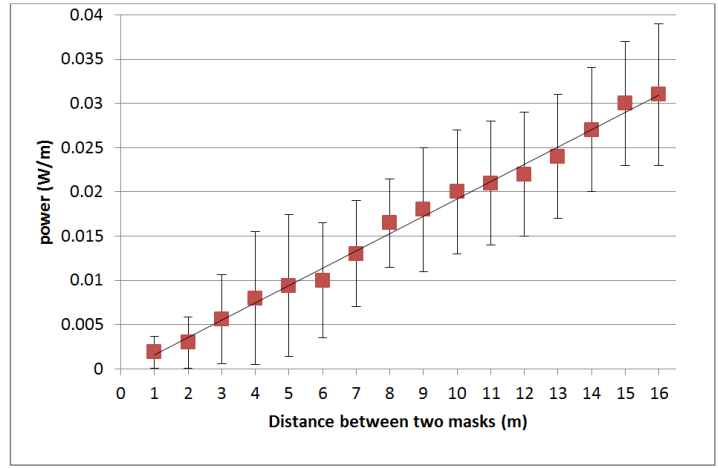

*Figure 7: Incident power per meter between two masks (PRELIMINARY)*

**Photon Energy Spectrum**

The influence of adding masks inside the undulator on the photon energy spectrum was studied. Figure 8 shows the energy spectrum when electron beam with128 GeV passes through an undulator module with and without the mask. It is considered that the circular aperture of the undulator module is 5.8mm and the mask aperture is 4.4mm respectively.

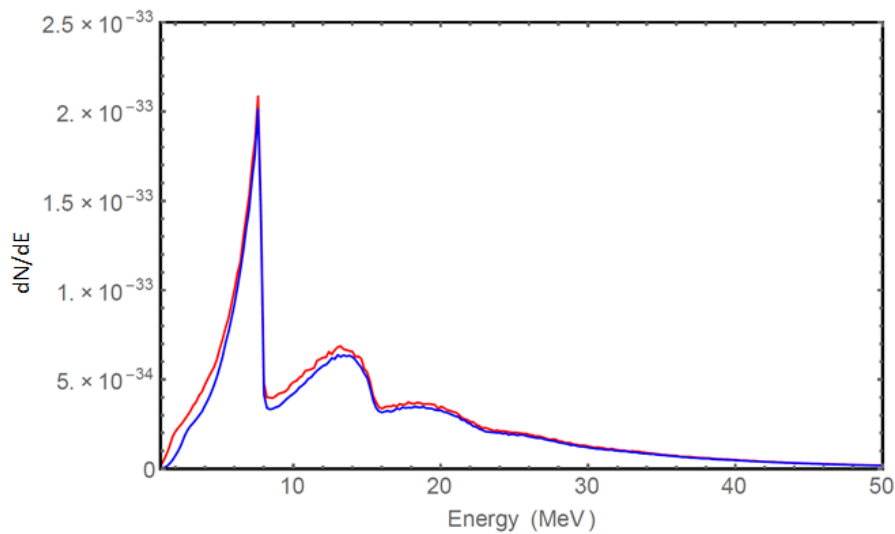

*Figure 8: the photon energy spectrum produced when a single electron with 128 GeV passes through a realistic helical undulator.*

The red line in the figure 8 shows the energy spectrum from the undulator module with a 5.8 mm aperture without the mask. On the other hand, the blue line represents the energy spectrum from the undulator with a mask of 4.4 mm aperture. Because low energy photons with large angles are more likely to be incident on the undulator wall [2], it is clear that masks with a 4.4 mm aperture can absorb photons with energy less than 1MeV.

**Photon Polarisation**

Besides the machine protection, a collimation system still serves other critical roles in physics. Principally, the ILC is used to increase positron polarisation [2]. Because circular photon polarisation created by an undulator differs with the emission angle, photon polarisation is likely to improve by addition of masks inside undulator [2]. It is known that positrons produced by polarised photon will inherit the photon polarisation. Therefore, increasing photon polarisation will automatically enhance positron polarisation.

The equation below was used to calculate photon polarization, $P_\gamma$ [1]:

$$P_Y = \frac{P_L - P_R}{P_L + P_R}$$

Whereby, $P_R$ and $P_L$ are the number of right and left handed photons. , $P_\gamma$ is the photon polarization as a function of photon energy.

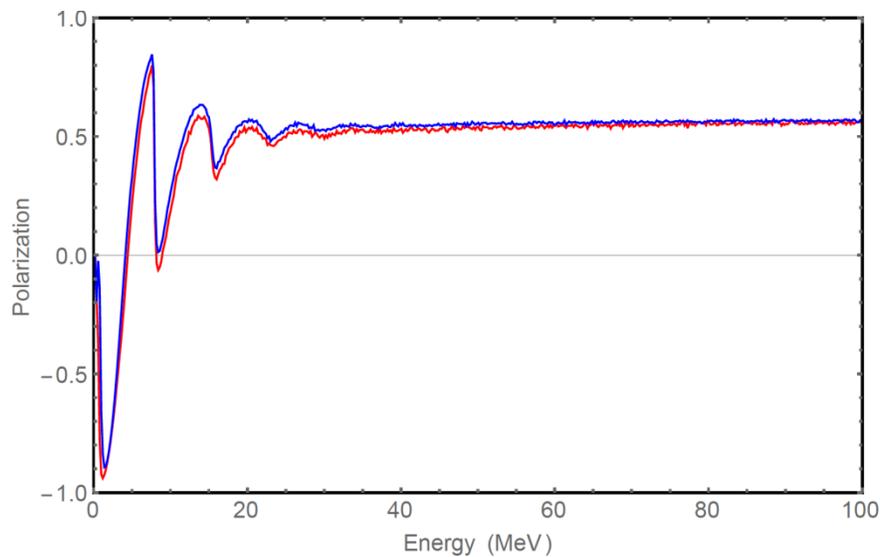

*Figure 9: shows photon polarisation as a function of the photon energy created by electron beam with 128-GeV energy.*

From the figure 9, the red line represents photon polarisation from the undulator without a mask. The blue line represents photon polarisation produced by the undulator with a mask having an aperture of 4.4 mm. This is an example of photon polarisation when electrons pass the 1.75 m long non-ideal undulator module. It is clear that adding masks inside the undulator increases photon polarisation.

**Conclusion**

In this paper, the power deposition on the undulator wall was considered. Because HUSR needs at least 86 m distance between the module and the observer to provide more accurate results, the incident power caused by $1^{st}$, $16^{th}$ and $33^{rd}$ modules was calculated from 88m range to the end of the undulator. It was discovered that the $1^{st}$ module could deposit preliminarily about 2.9 ±0.3 W/m at the end of the undulator. Since the maximum allowable heat load in the Superconducting undulator is 1 W/m, it was necessary to install masks inside the undulator to protect the undulator walls because they can absorb some of this power. Also, the incident power between masks was studied. However, the errors of calculation were high, masks could absorb this power. Lastly, the influence of placing masks inside the undulator on the photon energy spectrum was studied. It was known that Photons with low energy are more likely to hit the wall, those photons with energy lower than 1 MeV were absorbed by masks. The most important finding is that using masks can increase positron polarisation.